\documentclass[12pt,preprint]{aastex}

\shorttitle{sympathetic eruption, filament, and cause link}
\shortauthors{Song et al.}

\begin{document}

\title{Sympathetic eruptions of two filaments with an identifiable causal link observed by the Solar Dynamics Observatory}


\author{Zhiping Song\altaffilmark{1,2}, Yijun Hou\altaffilmark{2,3}, Jun Zhang\altaffilmark{1,2}, and Peng Wang\altaffilmark{1}}

\altaffiltext{1}{School of Physics and Materials Science, Anhui University, Hefei 230601, China; zpsong@ahu.edu.cn}

\altaffiltext{2}{CAS Key Laboratory of Solar Activity, National Astronomical Observatories,
Chinese Academy of Sciences, Beijing 100101, China; yijunhou@nao.cas.cn}

\altaffiltext{3}{University of Chinese Academy of Sciences, Beijing 100049, China}

\begin{abstract}
Filament eruptions occurring at different places within a relatively short time internal, but with a certain physical causal connection are usually known as sympathetic eruption. \textbf{Studies on sympathetic eruptions are not uncommon. However, in the existed reports, the causal links between sympathetic eruptions remain rather speculative.} In this work, we present detailed observations of a sympathetic filament eruption event, where an \textbf{identifiable} causal link between two eruptive filaments is observed. On 2015 November 15, two filaments \textbf{(F1 in the north and F2 in the south)} were located at the southwestern quadrant of solar disk. The main axes of them were almost parallel to each other. Around 22:20 UT, F1 began to erupt, forming two flare ribbons. The southwestern ribbon \textbf{apparently moved to} southwest and intruded southeast part of F2. This continuous intrusion caused F2's eventual eruption. Accompanying the eruption of F2, flare ribbons and post-flare loops appeared in northwest region of F2. Meanwhile, neither flare ribbons nor post-flare loops could be observed in southeastern area of F2. \textbf{In addition, the nonlinear force-free field (NLFFF) extrapolations show that the magnetic fields above F2 in the southeast region are much weaker than that in the northwest region.} These results imply that the overlying magnetic fields of F2 were not uniform. So we propose that the southwest ribbon formed by eruptive F1 invaded \textbf{F2 from its southeast region with relatively weaker overlying magnetic fields in comparison with its northwest region, disturbing F2 and leading F2 to erupt eventually.}

\end{abstract}


\section{Introduction}

Solar filaments are cool and dense plasma suspended in the corona, which are also known as prominences when observed
at the limb of the Sun. It is widely accepted that filaments are always located above the magnetic polarity inversion
lines (PILs). According to their locations, the filaments could be divided into three classes: active region filaments
forming inside active regions, intermediate filaments arising between active regions, and quiescent filaments which
are located at solar quiescent regions (Zirker et al. 1997; Martin 1998; Mackay et al. 2010). Filament eruptions,
solar flares, and coronal mass ejections are often seen as different manifestations of the same physical processes,
and filament eruptions are believed to play a key role in the onset of these eruptions (Schmieder et al. 2013;
Filippov 2018; Sinha et al. 2019). The popular filament eruption model is that a lifting filament stretches its
overlying magnetic field lines, creating a current sheet between the anti-parallel field lines, where magnetic
reconnection occurs, and then a bulk of plasma and magnetic structure are ejected into the interplanetary space.
During a filament eruption, the footprints of the reconnecting overlying magnetic field lines continuously brighten
\textbf{different regions in the chromosphere} and are observed as two flare ribbons which spread to both sides of
the filament \textbf{in a way of apparent motion}, while the post-flare loops just are the newly formed low-lying loop
lines (Kopp, \& Pneuman 1976; \textbf{Miklenic et al. 2007;} Shibata, \& Magara 2011; Hou et al. 2016; Yang \& Zhang 2018).

The eruptions of stable filaments are triggered by the loss of the balance of forces acting on them (Porfir'eva \& Yakunina 2013;
Kliem et al. 2014; Zaitsev \& Stepanov 2018). Several different mechanisms have been proposed and studied for the initiation
of filament eruption, such as flux emergence and cancellation (Zhang et al. 2001; Zheng et al. 2017; Dacie et al. 2018),
tether cutting reconnection (Moore et al. 2001; Cheng \& Ding 2016; Woods et al. 2018), breakout reconnection (Sterling et al.
2011; Kliem et al. 2013; Sun et al. 2015; Chen et al. 2016), and torus or kink magnetohydrodynamic (MHD) instability
(Aulanier et al. 2010; Riley et al. 2011; Hassanin \& Kliem 2016; Dechev et al. 2018; Hou et al. 2018; Duchlev et al. 2019).
A stable filament can also erupt due to the direct interaction from a surrounding filament, during which the magnetic
system of the original stable filament is disturbed by magnetic reconnections. Results from Yang et al. (2017) showed
that two nearby filaments, which were almost perpendicular to each other, could gradually approach to each other and
eventually erupt due to the direct collision between them. Furthermore, a filament could also erupt for the mass
injection from a surrounding erupting filament (Su et al. 2007).

In addition to the filament eruptions mentioned above, sympathetic eruptions between filaments are also ubiquitous.
These eruptions are defined as consecutive filament eruptions occurring within a short time in different locations
but having a certain physical connection (T{\"o}r{\"o}k et al. 2011; Jiang et al. 2014; Yang et al. 2012; Wang et al. 2016;
Joshi et al. 2016; Li et al. 2017). Two filaments, which reside above different PILs and share the
same overlying magnetic system, could interact sympathetically, i.e., the first erupting filament disturbs the
magnetic system of the second filament and leads to its eruption (Shen et al. 2012). A series of erupting filaments
could be connected by magnetic separatrices or quasi-separatrix layers. These filaments could be disturbed by each
other and erupted sympathetically by a chain of magnetic reconnections (Schrijver \& Title 2011). However, due to
the lack of direct observational evidences, it has been debated whether the close temporal correlation between
sympathetic eruptions is purely coincidental, or causally linked. The exact mechanism of sympathetic eruptions is still
not well understood (Biesecker \& Thompson 2000; T{\"o}r{\"o}k et al. 2011; Wang et al. 2018).

In the present work, we show a detailed process of sympathetic eruptions of two filaments by using high-quality data
from the Solar Dynamics Observatory (SDO; Pesnell et al. 2012). In this event, a flare ribbon caused by an eruptive filament
intruded the location of an adjacent filament and finally led it to erupt. The remainder of this paper is organized
as follows. Section 2 describes the observations and data analysis taken in our study. In Section 3, we investigate the
sympathetic event and present the results of observation in detail, which are followed by summary and discussion
in Section 4.

\section{Observations and Data Analysis}

We study sympathetic eruptions of two filaments in the southwest quadrant of solar disk on 2015 November 15-16.
The north filament (F1) lies in southwest of AR 12452 and the south filament (F2) is located at southeast of
AR 12449 and AR 12450. They are mainly observed by SDO/Atmospheric Imaging Assembly (AIA; Lemen et al. 2012)
and SDO/Helioseismic and Magnetic Imager (HMI; Schou et al. 2012). The AIA provides full-disk images taken
in 10 extreme ultraviolet (EUV) and ultraviolet (UV) wavelengths. The pixel
size and cadence of the EUV images are 0.{\arcsec}6 and 12s, (24 s for UV wavelengths), respectively. In this
study, images in five AIA EUV channels (131 {\AA}, 171 {\AA}, 193 {\AA}, 211 {\AA}, and 304 {\AA}) on 2015
November 15-16 are used for the multi-wavelength analysis of sympathetic eruptions of two filaments. We employ
the HMI line-of-sight (LOS) magnetic field data from the SDO to analyze evolution of the photospheric magnetic
fields related to the two filaments. These data have a pixel size of 0.{\arcsec}5 and a cadence of 45 s.
\textbf{The HMI and AIA images are preprocessed by standard routines in solar software package (SSW), and all
the data are differentially rotated to a reference time (00:00 UT on November 16).} To investigate the chromosphere
configuration of the associated filaments, we also employ corresponding
H$\alpha$ observations from the Global Oscillation Network Group (GONG; Harvey et al. 2011). GONG has collected H$\alpha$
images observed at seven sites around the world since mid-2010 and provides successive global H$\alpha$ observations
online.

\textbf{In order to reconstruct the three-dimensional (3D) magnetic fields above F1 and F2, we perform nonlinear force-free field (NLFFF)
extrapolation by using the optimization method (Wheatland et al. 2000; Wiegelmann 2004).
The boundary condition for the NLFFF extrapolation is an HMI full-disk vector magnetogram with a pixel spacing of 0.{\arcsec}5
at 22:00 UT on 2015 November 15, which is produced by using the Very Fast Inversion of the Stokes Vector (VFISV) inversion code
every 12 minutes (Borrero et al. 2011). The azimuthal component of the vector field is processed by Minimum Energy
method to resolve the 180{\degr} ambiguity (Metcalf 1994; Leka et al. 2009). Then we transform the vector magnetic field and the
geometric mapping of the observed field in the image plane into the heliographic plane (Gary \& Hagyard 1990).
The NLFFF calculation is conducted within a box of $600\times552\times512$ uniformly spaced grid points with $dx=dy=dz=1.{\arcsec}0$.
Moreover, we calculate the decay index $n$ of the horizontal magnetic fields above F2, which is defined as $n(z) =-zdln(Bh)/dz$
(Kliem \& T{\"o}r{\"o}k 2006) and provides important information about the strapping fields stabilizing the filament.}

\section{Results}

The event analyzed here occurred on 2015 November 15-16. As shown in Figure 1, two \textbf{sinistral} filaments (F1 in the north and F2
in the south) were located at the southwest quadrant of solar disk. The filament F1 lay in the southwest of AR 12452
and the filament F2 was near southeast of AR 12449 and AR12450 on November 15. \textbf{In H$\alpha$ observations, the lengths of 
them are about 500 Mm and 250 Mm, respectively.} Two filaments both lay above the PILs
and their main axes were almost parallel to each other. The filament F1 and filament F2 were separated by negative
magnetic fields between them.

Since about 22:20 UT on November 15, the northern filament (F1) began to rise quickly and then erupted. After examining
magnetograms and 193 {\AA} images, we notice that magnetic flux cancellation occurred around barbs of F1 accompanied
by 193 {\AA} brightening before F1 eruption (see Figure 2 and online animation of this figure). The FOVs of panels (c1)-(c3)
and panels (d1)-(d3) in Figure 2 are outlined by the blue rectangles in panels (a) and (b), respectively. And four
sites of flux cancellation along the PIL of F1 are marked by the green boxes where the magnetic fields with opposite
polarities kept approaching to each other (see panel (b)). In these four box regions, the magnetograms are replaced
by the corresponding HMI magnetograms on November 15 around 20:26 UT, 21:20 UT, 21:56 UT, and 20:35 UT, respectively.
To investigate the temporal evolution of the flux cancellation and brightening in detail, we focus on the box
region ``2''. Combining the AIA 193 {\AA} images and HMI magnetograms, one can see that the flux cancellation is
accompanied by the EUV brightening. From 17:00 UT to 22:15 UT, the emission enhancement (see panels (c1)-(c3)) as
well as the flux cancellation (see panels (d1)-(d3)) continuously appeared in this box region. Along the red line ``A-B''
shown in panel (d2), we make a time-slice plot (see panel (e)) in HMI LOS magnetograms. Moreover, the 193 {\AA} light
curve of the area contoured by the green curve in panel (c3) is superimposed on the time-slice plot (see the green
curve in panel (e)). It is shown that two magnetic fields with opposite polarities kept approaching to each other
in the box region ``2'' from 20:00 UT to 24:00 UT. Meanwhile, the emission strength mentioned above continued to enhance
from 21:00 UT to 22:20 UT and reached a peak at about 22:20 UT (see the red dotted line in panel (e)).

The detailed process of F1 eruption is displayed in Figure 3. We focus on the northwest area of F1 (see the FOV marked
by the green square in panel (a1)), where eruption of F1 is the most significant, to show the F1 eruption in detail.
Since about 22:20 UT, F1 started to rise and then erupted northwestward at about 23:00 UT, accompanied by formations of
two flare ribbons (outlined by the green dotted curves in panel (a3)). \textbf{These two flare ribbons appeared on both
sides of the magnetic neutral line of F1 and then apparently separated from each other in directions
perpendicular to the magnetic neutral line of F1.} The northeastern ribbon spread northeastward and
the southwestern one spread southwestward. To study the kinematic evolution of F1, we made time slices from 304 {\AA},
171 {\AA}, and 131 {\AA} images along the cut ``C-D'' shown in panel (a2). As displayed by the time-slice plots in
panels (b1)-(b3), the filament F1 \textbf{underwent three kinematic phases: slow rising, accelerating and constant velocity.}
By linear fitting, the erupting speeds of
F1 during \textbf{constant velocity} phase are estimated as about 100 km s$^{-1}$. The light curve of 193 {\AA} in the area
contoured by the green curve in Figure 2(c3) is superimposed on the time-slice image of 304 {\AA} (see the green curve in
panel (b1)). It can be seen that the peak time of brightening and eruptive time of F1 was almost coincident (see the vertical
white dotted line in panel (b1)). The detailed process of F1 eruption is also seen in online animation of Figure 3.

Once formation, \textbf{the southwestern flare ribbon related to eruptive F1 (FRF1)} continuously \textbf{apparently} moved
to southwest. \textbf{At about 23:50 UT, the northwestern part of FRF1 ceased its motion, but
the southeastern part of FRF1 kept moving southwestward, approaching the site of F2} and intruded the location of
F2 around 00:18 UT on November 16 (see Figure 4(a1)-(b1)). This intrusion lasted for
about 100 minutes. After that, the filament F2 began to lift slowly, and accelerated to erupt eventually (see panels (b2)-(b3)).
In order to investigate the kinematic evolution of F2 in detail, we made time slices (see panels (c1)-(c3)) from 304 {\AA},
171 {\AA} and, 131 {\AA} images along a slit ``E-F'' (see panel (b2)). In panels (c1)-(c3), the dashed curves
approximate the tracks of F2 in different wavelengths, the two vertical green dashed lines indicate the times when F2 began
to rise (t1) and to erupt (t2) respectively. We notice that F2 appeared as dark belts and it was almost stable before 00:20 UT,
when FRF1 reached the site of F2 (panel (b1), see also the green vertical dashed line ``t1''
in panels (c1)-(c3)). During 00:20 UT to 01:50 UT, F2 rose slowly accompanied the successive intrusion of FRF1 and the rising
speed was about 3 km s$^{-1}$. At about 01:50 UT, F2 began to rise quickly and then erupted (see the green vertical dashed
line ``t2'' in panels (c1)-(c3)), with the speeds of around 120 km s$^{-1}$. The temporal evolution of FRF1 spreading,
interaction between FRF1 and F2, and F2 eruption is also shown in online animation of Figure 4.

Since about 02:10 UT, the eruptive F2 also produced two flare ribbons at its northwest area (see the white solid squares
in Figure 5), which \textbf{appeared on both sides of the magnetic neutral line of F2 and then apparently separated from each
other in directions perpendicular to the magnetic neutral line of F2} (see panels (a1)-(a3)). The northern ribbon (FRF2) related
to F2 eruption approached FRF1 formed by F1 eruption. Finally, there displayed a clear separating line between the two
approaching ribbons (panel (a3)), implying that there is a magnetic separatrice between the two filaments and that F1
and F2 belong to two different magnetic systems. Seen from 171 {\AA}, 131 {\AA}, and 211 {\AA} images, two sets of
post-flare loops appeared in the northwest area in gradual phases after the eruptions of F1 and F2. The footpoints
of post-flare loops could be identified as the flare ribbons related to the eruptions of F1 and F2, respectively.
\textbf{There are two evidences that can be seen about the correlation between the flare ribbons (post-flare loops)
and their corresponding eruptive filaments. The first one is temporal correlation, i.e., FRF1 appeared shortly after the F1
eruption (see Figure 3), and FRF2 appeared following immediately the eruptive F2 (see Figure 4). The other one is spatial
correlation, i.e., the flare ribbons (post-flare loops) formed by different eruptive filaments appeared on both sides of
different magnetic neutral lines (see Figure 5).}
These post-flare loops and their footpoints are displayed in panels (b1)-(b3). It is shown that these loops were
rooted on the positive and negative magnetic fields on each side of the two neutral lines. However, we also notice
that neither clear flare ribbons nor obvious post-flare loops appeared in southeast region of F2 (see the dash-dotted
squares in Figure 5). The distribution of \textbf{flare ribbons} and post-flare loops related to eruptive F2 implies
that the overlying magnetic fields of F2 are not uniform, and they are weaker in southeast area but stronger in northwest
region. The temporal evolution of FRF2 and post-flare loops after F2 eruption is also shown in online animation of Figure 5.

\textbf{In order to confirm the non-uniformity of the magnetic fields above F2, We performed NLFFF
reconstruction of the overlying magnetic fields of F2 based on the photospheric vector magnetic fields at 22:00 UT
on 2015 November 15. Moreover, we calculated the decay index distribution of the reconstructed horizontal magnetic fields
above F2. As shown in Figure 6(a), the red-blue curve denotes the magnetic neutral line related to F2, and the
blue and red segments correspond to the southeast and northwest regions of F2, respectively. It is obvious that
the magnetic fields above F2 in the southeast region are weaker than those in the northwest region (see the green curves).
The results shown in panels (b) and (c) also reveal that the decay index of horizontal magnetic fields above F2 in
the southeast region is significantly larger than that in the northwest region. This means that the overlying horizontal
magnetic fields of F2 decay faster with increasing altitude in southeast region than in northwest region. The NLFFF modeling
results shown here well support the observations and our speculation that the overlying magnetic fields of F2 are not uniform.}

\section{Summary and Discussion}

Using the high-quality data from the SDO, we present eruptive processes of two filaments (F1 and F2) that occurred on 2015
November 15-16. F1 and F2 were separated by negative magnetic fields and their main axes were almost parallel to each other.
On November 15, a sequence of emergence and cancellation of the magnetic flux appeared inside the F1 channel and then F1
erupted. The erupting F1 produced two flare ribbons. The southwestern ribbon spread southwestward and intruded the southeastern
region of F2, which lasted about 100 minutes, and then F2 erupted. The eruptive F2 produced also two flare ribbons in its
northwestern region, which moved northward and southward respectively. During the evolution of these ribbons, there displayed
a clear separating line between the two nearby ribbons related to F1 and F2 eruptions, implying that there is a magnetic
separatrice between the two filaments and that F1 and F2 belong to different magnetic systems. In gradual phases after the
eruptions of F2, post-flare loops appeared in northwest region of F2, but neither flare ribbons nor post-flare loops were
observed in southeast area of F2. \textbf{Furthermore, the NLFFF extrapolated 3D magnetic fields and decay index distribution
of the horizontal fields above F2 reveal that the overlying magnetic fields of F2 in the southeast region are weaker than those
in the northwest region and that the overlying horizontal magnetic fields of F2 decay faster with increasing altitude in
southeast region than in northwest region.}

The eruptions of filaments are closely connected to the evolution of underneath magnetic fields, such as magnetic flux
emergence and cancellation. The relationship between them has been analyzed in previous studies (Mandrini et al. 2014;
Panesar et al. 2016; Zheng et al. 2017; Yang, \& Chen 2019). Zhang et al. (2001) discovered that when the major solar
eruption event (a giant filament eruption, a great flare, or an extended Earth-directed coronal mass ejection) happened,
the only obvious magnetic change in the course of the event is magnetic flux cancellation at many sites in the vicinity
of the filament. Li et al. (2015) found that the location of emerging flux within the filament channel is probably
crucial to filament eruption. They suggested that if the flux emergence appeared nearby filament's barbs, flux cancellation
was prone to occur, which probably caused the filament eruption. In the present work, prior to the eruption of F1, obvious
magnetic flux cancellation and emission brightening could be observed in F1 channel (Figure 2). When the emission
reached the peak, F1 began to erupt (see the dotted vertical white line in Figure 3(b1)). So, we suggest that the
flux cancellation happening in F1 channel results in F1 eruption.

As mentioned above, a filament probably erupts due to the magnetic flux emergence and cancellation. If there is another
stable filament located just beside this erupting filament coincidentally, the interaction between these two filaments
usually is inevitable and the direct collision would result in the eruption of the second filament (Bone et al. 2009;
Zheng et al. 2017). However, different from the filament-filament interactions, the sympathetic filament eruptions occur
without direct touching between the bodies of filaments or material of filaments (Bumba \& Klvana 1993; Wang et al. 2001;
Lugaz et al. 2017). The exact nature of sympathetic eruptions is still a controversial issue. Observational investigations
(Jiang et al. 2011; Shen et al. 2012; Joshi et al. 2016; Wang et al. 2018) and numerical simulations (Ding et al. 2006;
T{\"o}r{\"o}k et al. 2011; Lynch \& Edmondson 2013) suggest that magnetic reconnections of large-scale magnetic fields,
which are induced indirectly or directly by distant or nearby eruptions, may be the crucial physical links between
sympathetic eruptions, for the reconnections could weaken and partially remove the overlying magnetic fields on the
other filaments, resulting in their destabilization and eruptions.

In this work, F2 began to erupt about 100 minutes after FRF1 intruded southeast
location of F2 (see Figures 4(b1)-(b3)). And then, we observed two different manifestations in northwest and southeast regions of F2.
There were clear flare ribbons formed by the eruptive F2 in the northwest area of F2, which spread northward and
southward respectively and the northern one extended to the edge of FRF1 (see Figure 5(a3)). Meanwhile, clear post-flare
loops were also observed in the northwest region of F2. But, neither clear flare ribbons nor obvious post-flare loops
could be observed in the southeast area of F2 (see Figure 5, and the online animation of Figure 5). These \textbf{observational}
results imply that the distribution of the overlying magnetic fields of F2 is not uniform, i.e., the overlying magnetic
fields of F2 are stronger in its northwest area, but weaker in its southeast region. \textbf{The NLFFF reconstructions
well support the observations and our speculation that the overlying magnetic fields of F2 are not uniform (see Figure 6).
As shown in Figure 4 and its corresponding movie, the northwestern part of FRF1 ceased its apparent motion to the southwest
at about 23:50 UT on November 15, meanwhile the southwestern part of FRF1 kept moving southwestward, approaching the site of F2.
We suggest that it is possible that} the stronger overlying magnetic fields of F2 in northwest area partially blocked FRF1
from expanding southwestward further. \textbf{In contrast,} the weaker overlying magnetic fields in the southeast region of
F2 made F2 to be more sensitive
to external disturbances happening in that region. As a result, \textbf{when the southeast part of FRF1 continuously
intruded the southeast site of F2, F2 was disturbed to slowly rise and erupt eventually. Note that the AIA observations of full
eruption of F2 shows that no active eruption is clearly observed
in the southeast part of F2 at the initial phase. The possible reasons, in our opinions, are as follows: Firstly, the speed at which F2
started to rise is too small to distinguish clearly. Secondly, as shown in the H$\alpha$ observations, the plasma density
of F2 in the southeast region is much less than that in the northwest region (see Figure 1(a)). So it is reasonable
that the eruption of F2 appeared more obviously in the northwest region even though it started in the southeast region.
Finally, the projection effect may also make a difference here, causing the observations different from the real situations.}

Because no any direct touch between F1 and F2 was observed, we conjecture that the interaction occurring between
F1 and F2 belongs to sympathetic one. And the causal link of the sympathetic interaction is that FRF1 invades
the southeastern region of F2, where the overflying magnetic fields of F2 are very weak. Based on the multi-wavelength
observations, we present a schematic diagram in \textbf{Figure 7} to illustrate this sympathetic eruptions event. Difference
of field line density in \textbf{Figure 7} is used to indicate the non-uniformity distribution of the overlying magnetic
fields of F2. Panel (a) represents the initial magnetic topology of the two filaments and the overlying ambient
coronal fields. Flux cancellation happening around the barbs of F1 disturbs F1 (see Figure 2) and results in
its final eruption (see Figure 3). Then two flare ribbons are formed after the eruption of F1, one of which keeps
spreading toward F2, intruding and disturbing the magnetic system of F2 in its southeast area (panel (b)). As
the overlying magnetic fields of F2 are very weak in the disturbed location, the successive disturbance destabilizes
F2, and F2 erupts eventually (panel (c), also see Figure 4). After the eruptions of F2, the clear flare ribbons
(shown by the thick purple curves) and post-flare loops (shown by the thick red curves in panel (d)) only appear
in F2's northwest region, where the overlying magnetic fields are strong (panels (c)-(d), also see solid squares
in Figure 5). And the thin purple and red curves in panels (c)-(d), respectively, are employed to indicate the
ambiguous flare ribbons and post-flare loops in F2's southeast area, where the overlying magnetic fields are very
weak (also see the dash-dotted squares in Figure 5).

\textbf{In the existing research reports on sympathetic filament eruptions, the directly observable criterion is
mostly the correlation of the eruptive time between two filaments}. Joshi et al. (2016) interpreted two successive
eruptions as sympathetic eruptions, which occurred in different active regions and associated by a chain of
magnetic reconnections. The phenomenon that fast rise of the second filament began soon after the first filament
eruption was suggested as sympathetic eruptions (Li et al. 2017). In the present work, in addition to the temporal
correlation, an \textbf{identifiable} causal link between two sympathetic eruptive filaments was observed.
To our knowledge, similar observations have never been reported before.

\acknowledgments{
SDO is a mission of NASA's Living With a Star (LWS) Program. AIA and HMI data are courtesy of the NASA/SDO science
teams. This work is supported by the National Natural Science Foundations of China (U1531113, 11903050, 11533008,
11790304, 11773039, 11673035, 11673034, 11873059 and 11790300), the open topic of the Key Laboratory of Solar Activities
of the Chinese Academy of Sciences (KLSA201902), the NAOC Nebula Talents Program, and Key Programs of the Chinese Academy of Sciences (QYZDJ-SSW-SLH050).
}

\clearpage

\begin{figure}
\centering
\includegraphics [width=0.93\textwidth]{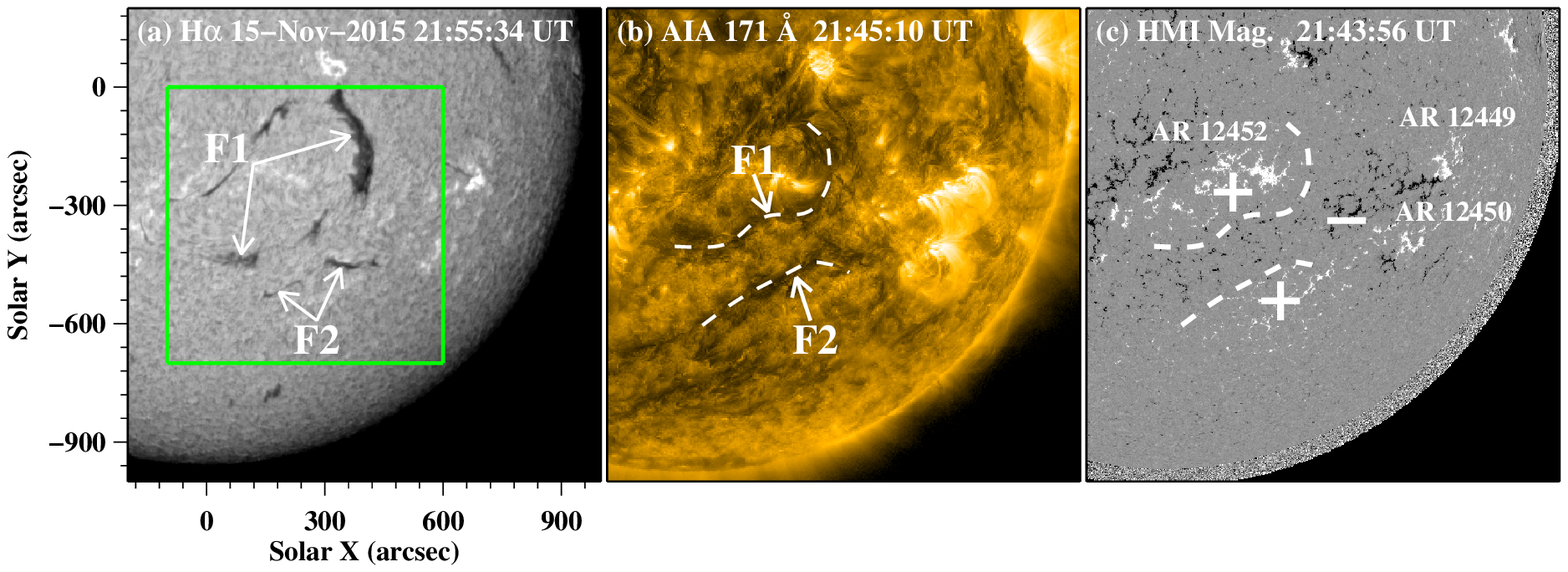}
\caption{Overview of the two filaments on 2015 November 15. Panel (a): H$\alpha$ image. The green square in panel (a) outlines the FOVs of panels
(a)-(b) in Figure 2, panels (a1)-(a3) and (b1)-(b3) in Figure 4 and, Figure 5. Panel (b): AIA 171 {\AA} image. Panel (c): HMI LOS magnetogram. The
two white dashed curves in panels (b)-(c) show the approximate locations of the two nearby filaments before eruptions.
}
\label{fig1}
\end{figure}

\begin{figure*}
\centering
\includegraphics [width=0.75\textwidth]{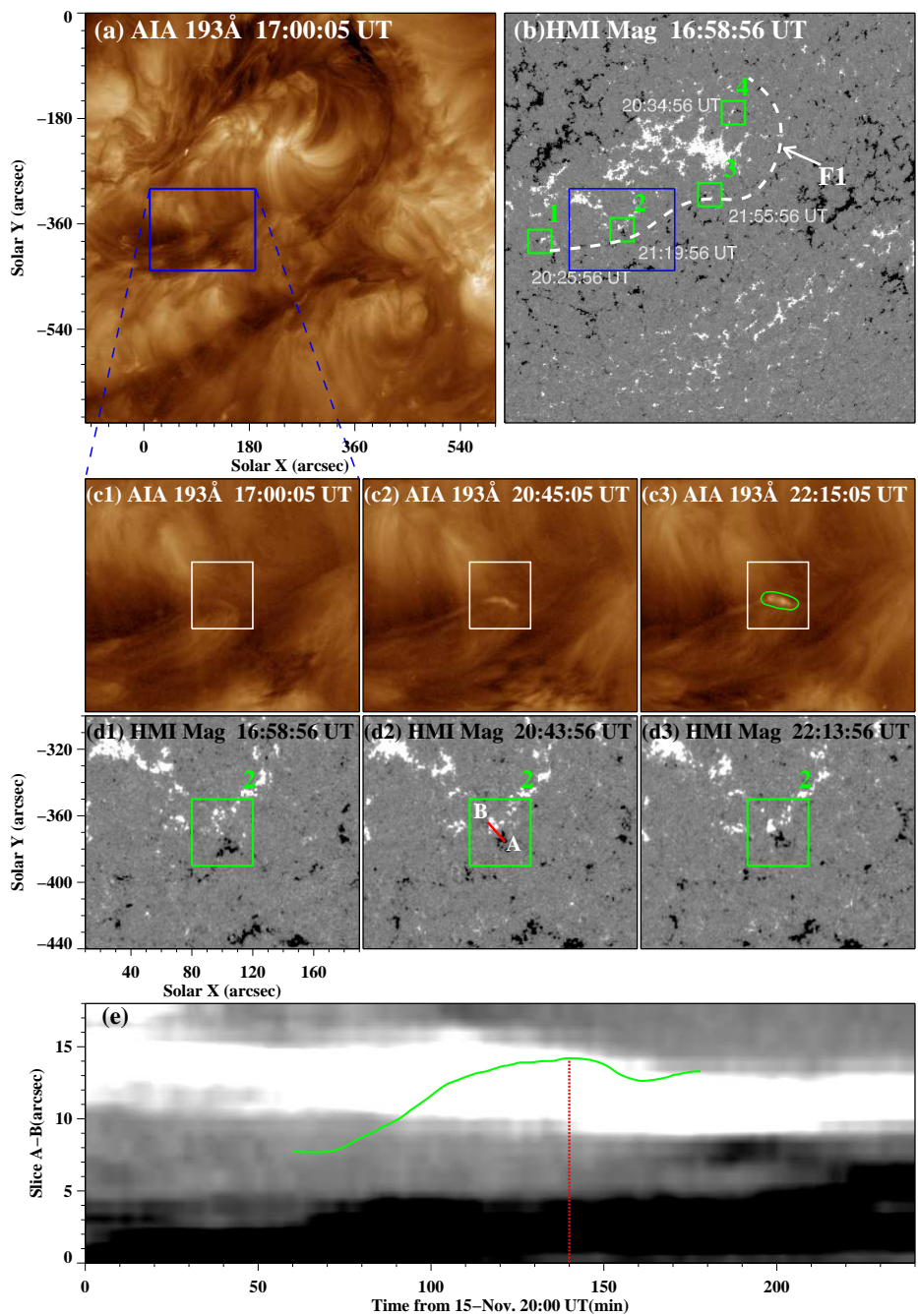}
\caption{AIA 193 {\AA} images and HMI LOS magnetograms showing the emission brightening and flux cancellation near the barbs of F1. Panels (a)
and (b): 193 {\AA} image and HMI magnetogram showing the sites of brightening and flux cancellation. Panels (c1)-(d3): Time sequences of 193 {\AA}
images and HMI magnetograms showing the temporal evolution of the brightening and flux cancellation, respectively. The white and green squares
correspond to the green box ``2'' in panel (b). The green curve in panel (c3) is the contour of emission strength in 193 {\AA} wavelength. Panel
(e): Time-slice plot of magnetograms along the cut ``A-B'' marked in panel (d2). The animation shows the brightening and flux cancellation evolving
from 17:00 to 22:50 UT on 2015 November 15.
}
\label{fig2}
\end{figure*}

\begin{figure*}
\centering
\includegraphics [width=0.90\textwidth]{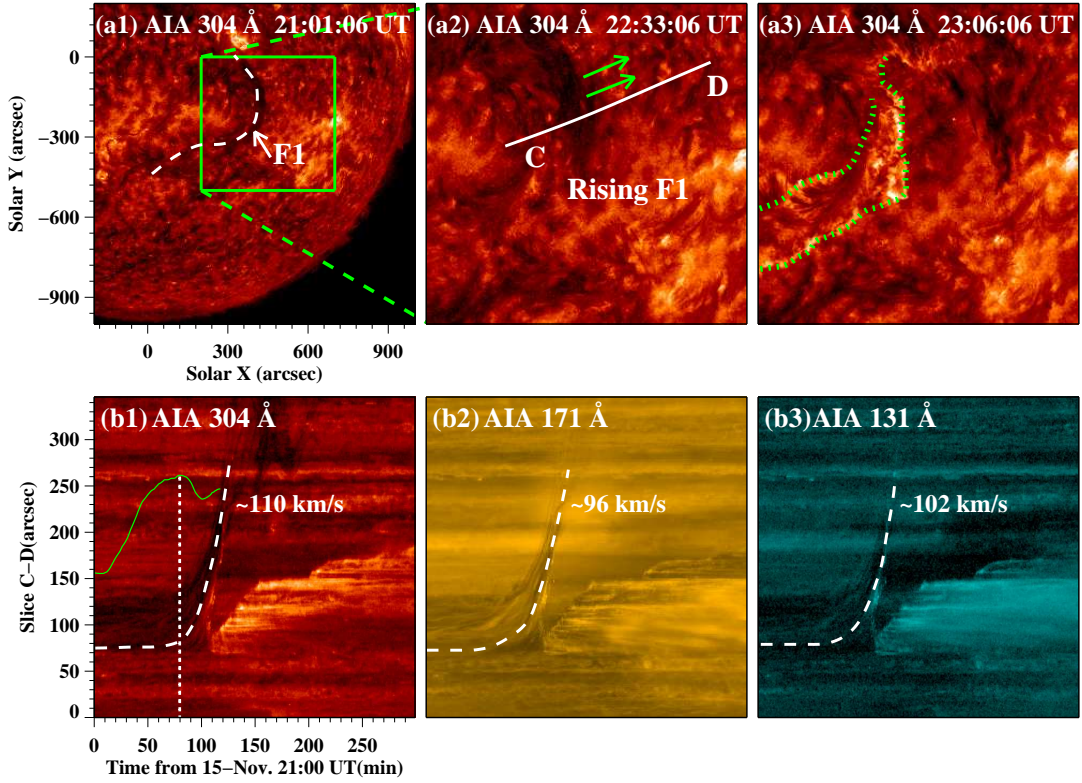}
\caption{Images of AIA 304 {\AA} and time-slice plots showing the eruption of F1. Panels (a1)-(a3): Time sequences of 304 {\AA} images. The green
arrows in panel (a2) denote the projected moving directions of F1. The green dotted curves in panel (a3) exhibit two flare ribbons caused by F1
eruption. The line ``C-D'' in panel (a2) marks the position of time-slice plots in panels (b1)-(b3). Panels (b1)-(b3): Time-slice plots of 304 {\AA},
171 {\AA}, and 131 {\AA} images displaying the temporal evolution of F1. The dashed curves approximate the tracks of F1 in different wavelength images. The green curve in panel (b1) is the 193 {\AA} light curve in the region marked by the green curve in Figure 2(c3). The vertical white dotted line in panel (b1) indicates the time of emission reaching at the peak (the onset of F1 eruption). The animation shows the eruption process of F1 from 22:10 to 23:28 UT on 2105 November 15.}
\label{fig3}
\end{figure*}

\begin{figure*}
\centering
\includegraphics [width=0.90\textwidth]{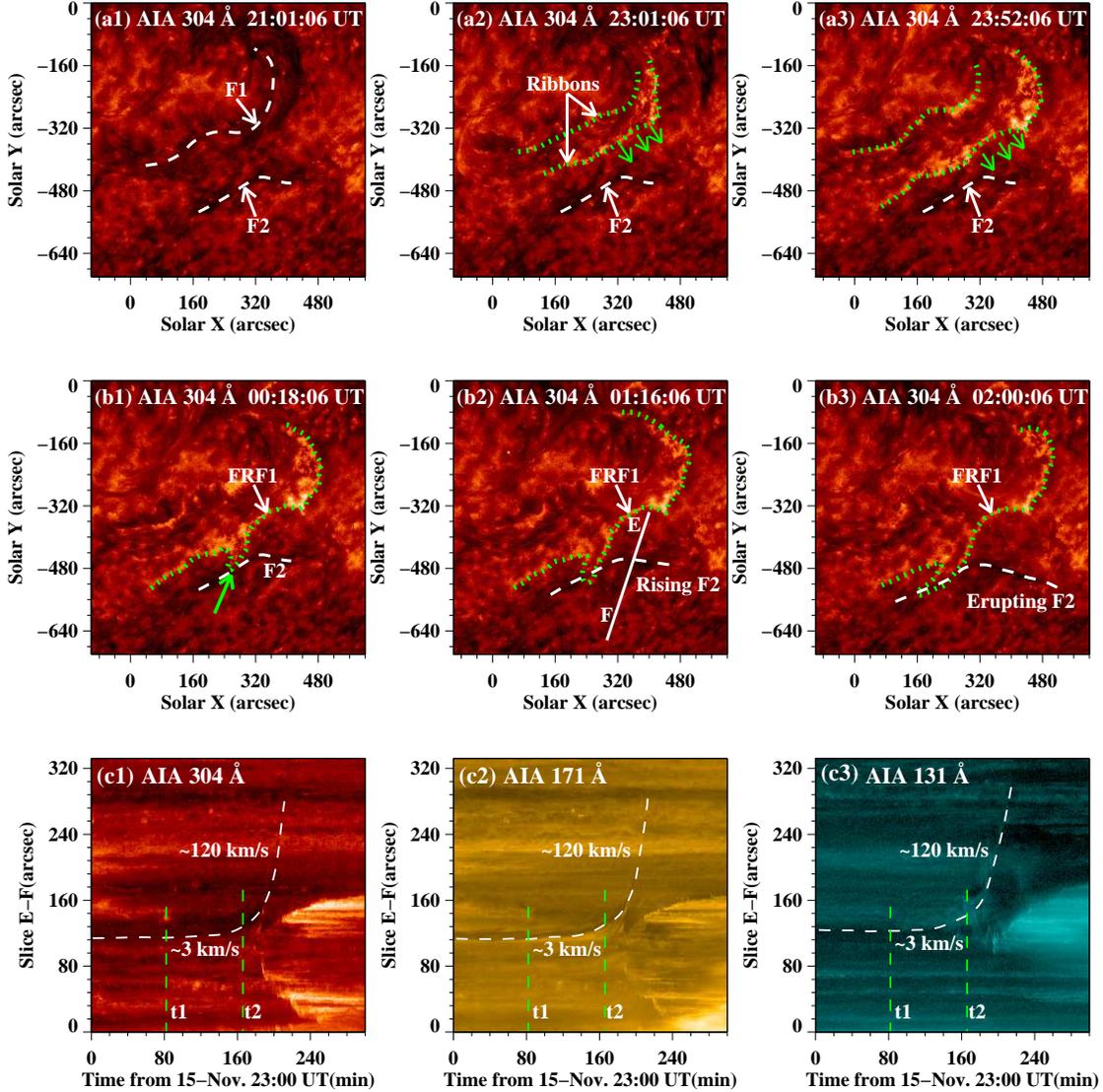}
\caption{The temporal evolution of FRF1 spreading, interaction between FRF1 and F2, and F2 eruption. Panels (a1)-(a3): Time sequences of 304 {\AA} images showing the expansion of the flare ribbons formed by F1 eruption. In panel (a1), the
approximate locations of F1 and F2 before their eruptions are outlined by white dashed curves. The green dotted curves in panels (a2)-(a3) display
the separating flare ribbons related to F1 eruption. The green arrows in panels (a2)-(a3) indicate the spreading direction of the southwestern ribbon.
Panels (b1)-(b3): Time sequences of 304 {\AA} images displaying the lifting and eruption of F2 by the intrusion of the southwestern flare ribbon (FRF1) related to erupting F1. The green arrow in panel (b1) denotes the location where FRF1 reached and intruded F2. The line ``E-F'' in panel (b2) marks the position of time-slice plots in panels (c1)-(c3). Panels (c1)-(c3): time slices of AIA 304{\AA}, 171 {\AA}, and 131 {\AA} images revealing the temporal evolution of F2. The animation shows the spreading of FRF1, interaction between FRF1 and F2, and F2 eruption from 00:00 to 02:00 UT on 2015 November 16.}
\label{fig4}
\end{figure*}

\begin{figure*}
\centering
\includegraphics [width=0.90\textwidth]{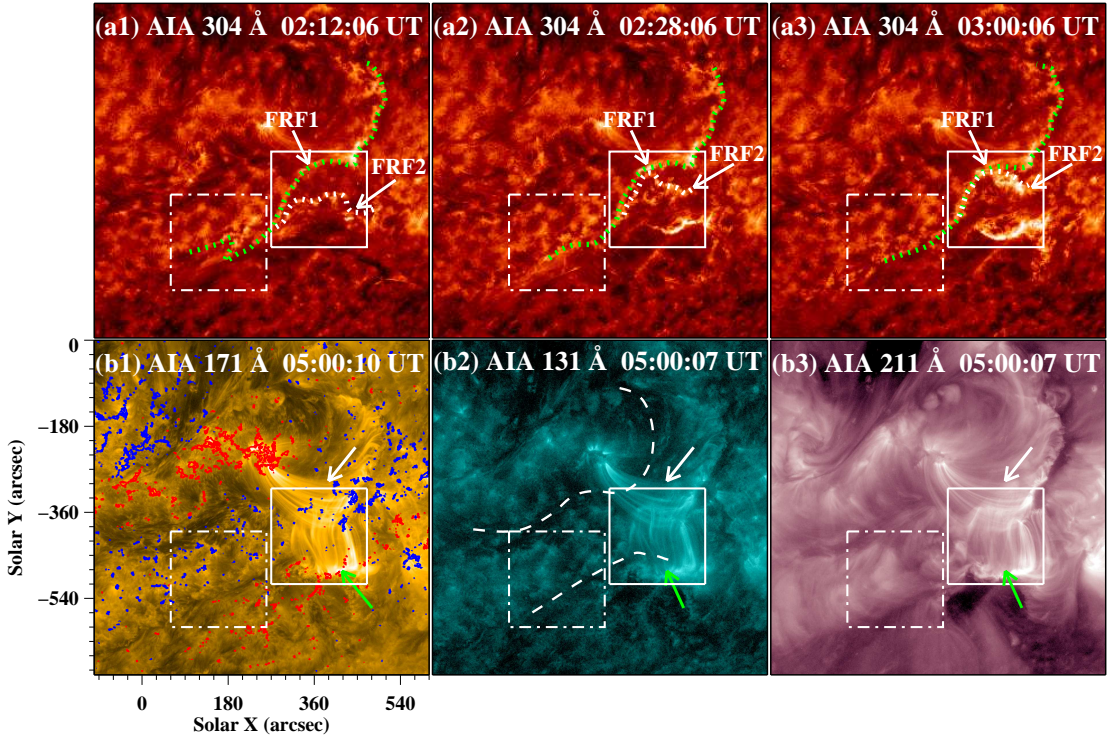}
\caption{The temporal evolution of FRF2 and post-flare loops after F2 eruption. Panels (a1)-(a3): Time sequences of AIA 304 {\AA} images showing the process of the northern ribbon (FRF2; white dotted curves) related
to F2 eruption approaching the southwestern ribbon (FRF1; green dotted curves) formed by F1 eruption. Panels (b1)-(b3): AIA 171 {\AA}, 131 {\AA},
and 211 {\AA} images displaying the post-flare loops caused by the eruptions of F1 and F2. The red and blue curves in panel (b1) are contours of
corresponding HMI LOS magnetic fields with positive and negative polarity, respectively. The contour levels are $\pm$50 Gauss. The white and
green arrows in panels (b1)-(b3) indicate the post-flare loops formed after F1 and F2 eruptions, respectively. The dashed curves in panel (b2)
denote the approximate locations of F1 and F2 before their eruptions. The white and dash-dotted squares in all panels outline two different regions
of northwest and southeast of F2, respectively. The animation shows the formation and spreading of FRF2, and the post-flare loops caused by the
eruptions of F1 and F2 from 02:00 to 04:59 UT on 2015 November 16.}
\label{fig5}
\end{figure*}

\begin{figure*}
\centering
\includegraphics [width=0.90\textwidth]{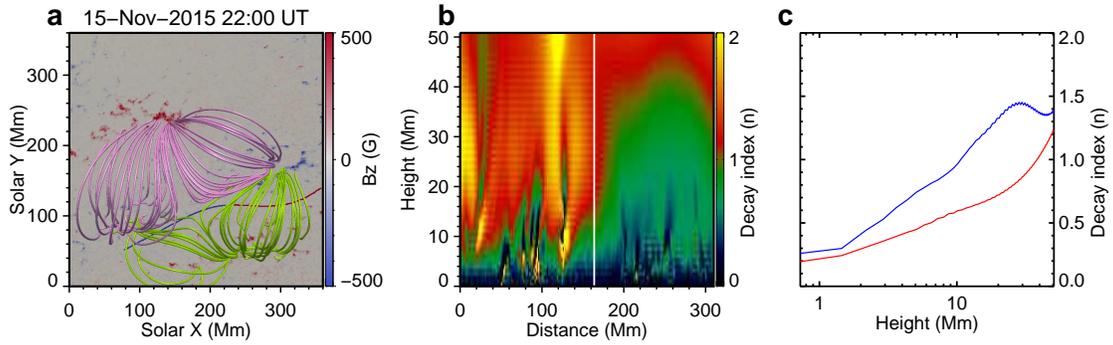}
\caption{\textbf{Results of NLFFF reconstruction of overlying magnetic fields of F1 and F2. Panel (a): Top view of the magnetic fields above F1 and F2. Pink and green curves represent the overlying magnetic field lines of F1 and F2, respectively. The blue-red curve marks the magnetic neutral line related to F2, and the blue and red segments correspond to the southeast and northwest regions of F2, respectively. Panel (b): The horizontal magnetic field decay index distribution in the vertical plane above the blue-red curve labeled in panel (a). Left (right) side of the white vertical line corresponds to the southeast (northwest) region of F2. Panel (c): The decay index of overlying horizontal magnetic fields of F2 as a function of height in two different regions. The blue curve indicates the southeast region, and the red one indicates northwest region.}}
\label{fig6}
\end{figure*}

\begin{figure*}
\centering
\includegraphics [width=0.90\textwidth]{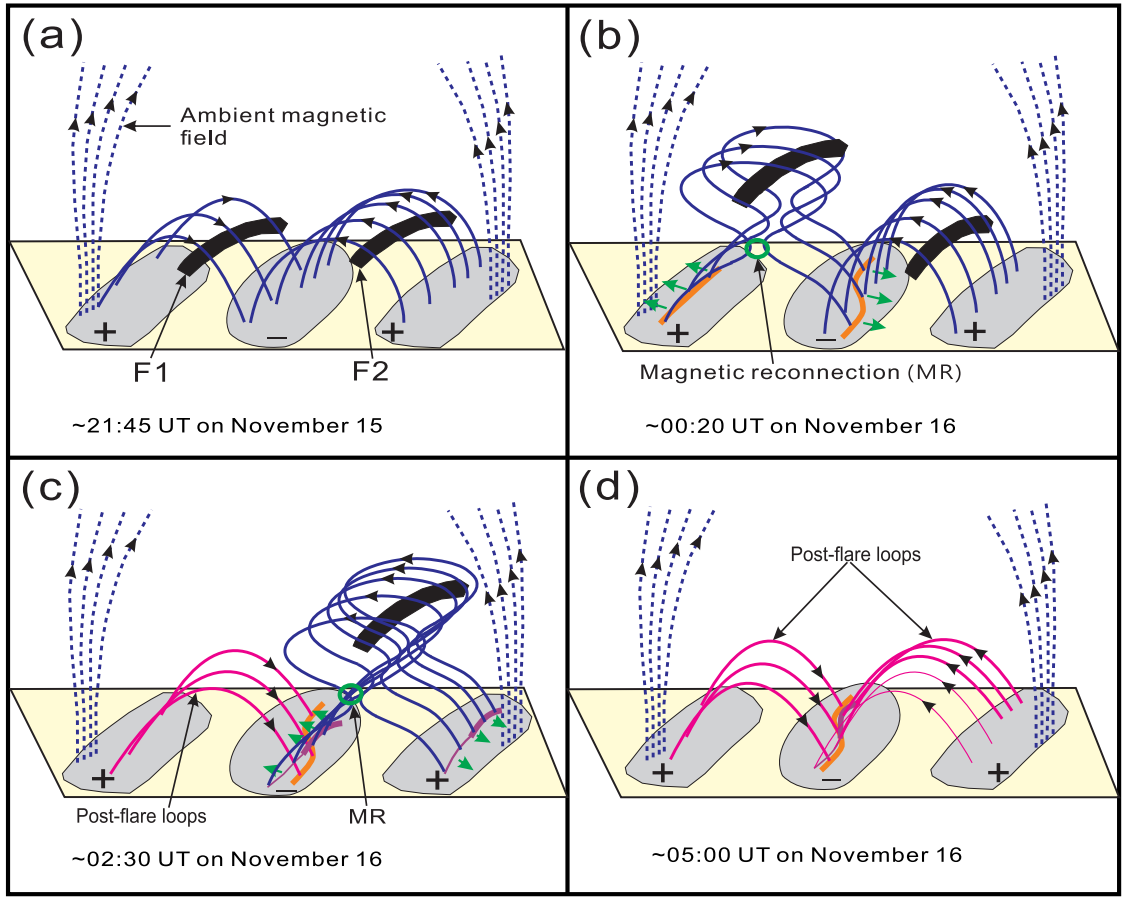}
\caption{Schematic diagram illustrating the sympathetic eruption event. Panel (a): The initial magnetic configuration of filaments F1 and F2.
Panel (b): The eruption of F1 and spreading of flare ribbons. The green arrows in panel (b) indicate the spreading direction of the flare
ribbons formed by F1 eruption. Panels (c): The eruption of F2 and spreading of flare ribbons. The green arrows in panel (c) show the spreading
direction of the flare ribbons related to the F2 eruption. Panel (d): Post-flare loops and approaching flare ribbons after the eruption of F1
and F2. In all panels, the blue dashed curves represent the ambient magnetic field lines. Orange curves in panels (b)-(d) show the flare
ribbons formed by F1 eruption. Purple curves in panels (c)-(d) indicate the flare ribbons related to F2 eruption. The green circles in
panels (b)-(c) display the sites of magnetic reconnection.}
\label{fig7}
\end{figure*}

\end{document}